\documentclass[12pt,preprint]{article}
\usepackage{amssymb}

\usepackage{amsmath}
\usepackage{graphics}
\usepackage{epsfig}
\usepackage{bbm}

\newcommand{\eqb}{\begin{equation}}
\newcommand{\eqe}{\end{equation}}
\newcommand{\dmb}{\begin{displaymath}}
\newcommand{\dme}{\end{displaymath}}

\newcommand{\eab}{\begin{eqnarray}}
\newcommand{\eae}{\end{eqnarray}}

\newcommand{\be}{\begin{equation}}
\newcommand{\ee}{\end{equation}}

\begin{document}

\begin{titlepage}
\begin{flushright} 
\end{flushright}
\vspace{0.6cm}

\begin{center}
\Large{Explosive Z Pinch}

\vspace{1.5cm}

\large{Francesco Giacosa$\mbox{}^*$, Ralf Hofmann$\mbox{}^\dagger$, and 
Markus Schwarz$\mbox{}^\dagger$}

\end{center}
\vspace{1.5cm} 

\begin{center}
{\em $\mbox{}^*$Institut f\"ur Theoretische Physik\\ 
Johann Wolfgang Goethe -- Universit\"at\\ 
Max von Laue -- Str. 1\\ 
D-60438 Frankfurt am Main, Germany\vspace{0.5cm}\\ 
$\mbox{}^\dagger$Institut f\"ur Theoretische Physik\\ 
Universit\"at Heidelberg\\ 
Philosophenweg 16\\ 
69120 Heidelberg, Germany
}
\end{center}
\vspace{1.5cm}
\begin{abstract}

We propose an explanation for the recently observed powerful 
contained explosion in a Z pinch experiment performed at Sandia 
National Laboratories. Our arguments are based on the 
assumption that a pure SU(2) Yang-Mills theory of 
scale $\sim 0.5\,$MeV is responsible for 
the emergence of the electron and its neutrino.   
    
\end{abstract} 

\end{titlepage}

\section{Introduction}

Very recently, an unexpected powerful contained explosion was detected in a
Z pinch experiment at Sandia National Laboratories. Preceding the explosion, 
an ion temperature $\sim$300\,keV was reached shortly after stagnation, and the energy
radiated away in soft X-rays was observed to be 3 to 4 times the estimated
kinetic energy released by the intersection of ions and electrons when
accelerated towards the plasma axis in the course of implosion. 
In \cite{Haines2006} a model was proposed which explains the rapid conversion of the
magnetic-field energy to thermal energy of ions. The conversion mechanism employs short wavelength $m=0$
magnetohydrodynamic (MHD) instabilities forming at stagnation
which lead to viscous ion (and subsequently electron) heating. In this way, the
observed imbalance between the kinetic energy of the implosion and of the energy radiated in 
soft X-ray emission was addressed. 

The purpose of this Letter is to point out a likely reason for the 
contained explosion observed in high-temperature 
Z pinches at Sandia \cite{Zmachine}.

\section{$SU(2)_e$ in its confining phase\label{conf}}

Our analysis is based on the postulate that the emergence of the electron
and its neutrino is due to pure SU(2) gauge dynamics subject to a Yang-Mills
scale $\Lambda\sim m_e$. A pure Yang-Mills theory is defined solely in terms
of gauge fields; no matter fields occur in the fundamental Lagrangian.
Thermodynamically, SU(2) Yang-Mills theory comes in three phases. In order
of decreasing temperature there is a deconfining, a preconfining, and a
confining phase \cite{Hofmann2005,Hofmann2006}. Here we are only interested in the
confining phase.

\begin{figure}[tbp]
\begin{center}
\leavevmode
\leavevmode
\vspace{4.3cm} \includegraphics{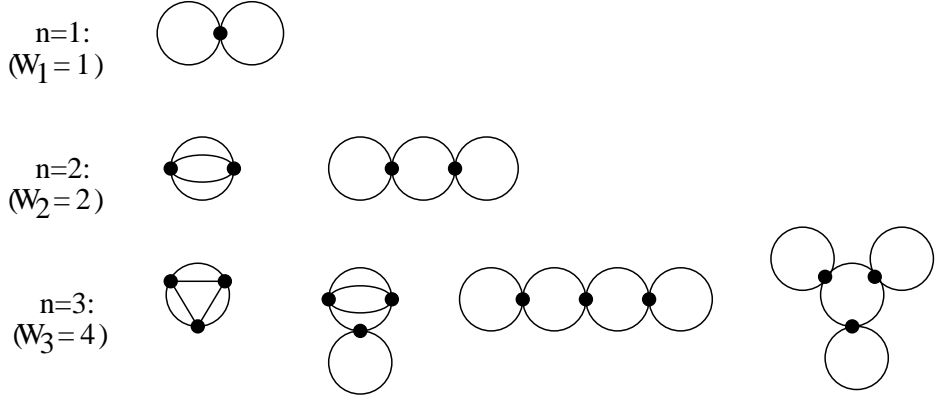}
\end{center}
\caption{{The topologies of the solitonic excitations for an SU(2)
Yang-Mills theory in the confining phase. Only the excitations with up to 
$n=3$ selfintersections are depicted.}}
\label{Fig-1}
\end{figure}

The excitations in that phase are single and selfintersecting center-vortex
loops \cite{Hofmann2005}. Each intersection point carries one unit of
electric charge (each sign is equally likely) and a mass given by $m_{e}$:
The mass spectrum thus is equidistant, $m_{n}=n\cdot m_{e}$. Two of the
four flux tubes connected to an intersection point exhibit the same flux direction; 
the other two have the opposite flux. All flux tubes are infinitely thin. In a
given flux-tube segment the direction of the flux is twofold degenerate: For a 
given flux-direction also the oppositely directed flux exists. 
Moreover, for a given soliton it is possible to go around the entire
flux-system along a closed curve. This is why we identify each soliton with a
spin-1/2 fermion. In the presence of propagating photons 
\cite{Hofmann2005,GiacosaHofmann2005,SchwarzGiacosaHofmann2006} the
only stable excitations are the single and the one-fold selfintersecting
center-vortex loop. The former is identified with the (electron-) neutrino
while the latter represents the electron or the positron. 
Fig.\thinspace \ref{Fig-1} shows the distinct topologies of selfintersecting
center-vortex loops up to intersection number $n=3$. There is a
charge-multiplicity factor $c_{n,k}\ge n+1$ associated with the $k$th 
soliton of $n$ selfintersections: Each intersection point may carry charge $+$ or $-$, and
we do not take into account any ordering of charges \footnote{%
Setting $c_{n,k}\equiv n+1$ for $n>2$ represents only a lower bound on
the actual charge multiplicity. For example, at $n=3$ the second and the third
topology in Fig.\thinspace \ref{Fig-1} have a charge multiplicity of $c_{3,2}=c_{3,3}=6$ 
instead of 4.}.

Modulo the charge multiplicities the number of distinct
solitons is given by the number of distinct topologies of connected vacuum diagrams
in a $\lambda \phi ^{4}$ quantum field theory. This theory was investigated
carefully in \cite{KleinertBook2002} and references therein. 
Up to $n=6$ the number $W_{n}$ of topologically distinct and connected vacuum diagrams in a 
$\lambda \phi ^{4}$-theory is given as follows \cite{KleinertBook2002}: 
\begin{equation}
W_{1}=1\,,\ \ \ W_{2}=2\,,\ \ \ W_{3}=4\,,\ \ \ W_{4}=10\,,\ \ \
W_{5}=28\,,\ \ \ W_{6}=97\,.  \label{vacdiagr}
\end{equation}%
Let us proceed by assuming that all solitons are absolutely stable: 
Long-range interactions between the charges, mediated by photons, are assumed to be 
switched off and solitons are separated sufficiently such that no contact
interactions can occur. The logarithm of the partition function $Z$ for the system of
massive spin-1/2 states then takes the following form 
\begin{equation}
\log Z=\sum_{n=0}^{\infty }\log Z_{n}\,,\ \ \ \ \ \ \ \ \log
Z_{n}=M_{n}V\int \frac{p^{2}dp\,}{2\pi ^{2}}\log \left( 1+e^{-\beta \omega
_{n}(p)}\right) \,  \label{partfunc}
\end{equation}%
where $\omega _{n}(p)=\sqrt{p^{2}+(n\cdot m_{e})^{2}}$, $V$ denotes the
(thermodynamically large) volume of the system, and $\beta\equiv 1/T_e$ is the inverse (electron)
temperature. The number $M_{n}$
represents the total multiplicity of solitons with $n$ selfintersections. It is given as 
\begin{equation}
M_{n}=2\times\sum_{k=1}^{W_{n}} c_{n,k}\ge 2\,W_n\,(n+1)  \label{mn}
\end{equation}%
where the factor $2$ takes into account the spin degeneracy. The total 
pressure $P$ and the total energy density $\rho$ are obtained from $Z$ as 
\begin{equation}
P=T\frac{\partial \log Z}{\partial V}=\sum_{n=1}^{\infty }P_{n}\,,\ \ \ \ \ \ \ \ \ 
\rho =T\frac{\partial P}{\partial T}-P\,=\sum_{n=1}^{\infty }\rho _{n}
\label{pressenerg}
\end{equation}%
where $P_{n}$ and $\rho _{n}$ refer to the pressure and energy density of particles 
with mass $m_{n}=n\cdot m_{e}$ (solitons with $n$ selfintersections). Explicitly, we have
\begin{equation}
P_{n}=M_{n}\,T\int_{0}^{\infty }\frac{p^{2}dp}{2\pi ^{2}}\,\ln [1+e^{-\beta
\omega _{n}(p)}]\,,\ \ \ \ \ \text{ }\rho _{n}=M_{n}\int_{0}^{\infty }\frac{p^{2}dp}{%
2\pi ^{2}}\frac{\omega _{n}(p)}{e^{\beta \omega _{n}(p)}+1}\,.  \label{rhon}
\end{equation}
For $n=1$ we recover the usual expression for an electron-positron gas with
chemical potential $\mu=0$ at temperature $T$. The case $n=0$ (neutrinos) is not 
important for the present discussion.

In \cite{BenderWu1967} a lower bound $W_{n,\tiny\mbox{low}}$ for $W_{n}$
was obtained: 
\begin{equation}
W_{n}>n!\,3^{-n}\equiv W_{n,\tiny\mbox{low}}\,.  \label{Wlowerbound}
\end{equation}%
Using $W_{n,\tiny\mbox{low}}$ and setting $c_{n,k}\equiv n+1$, one finds that 
$M_{n}$ is bounded from below by $M_{n,\tiny\mbox{low}}\equiv 2(n+1)!\,3^{-n}$. 
We use $M_{n,\tiny\mbox{low}}$ to obtain a lower estimate 
for the partial pressure $P_{n}$ at large $n$.
\begin{figure}[tbp]
\begin{center}
\leavevmode
\leavevmode
\vspace{4.3cm} \includegraphics{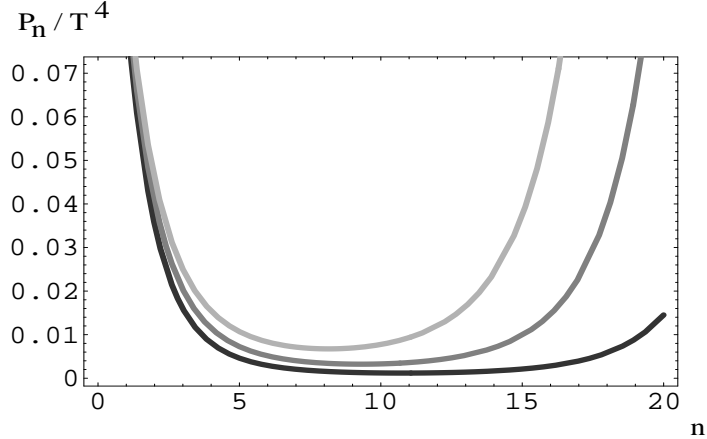}
\end{center}
\caption{{Lower bound for the ratio of 
partial pressure $P_n$ to $T^4$ as a function of $n$ for three different temperatures: 
$T=0.65\,m_e$ (dark grey),$T=0.7\,m_e$ (grey), and $T=0.75\,m_e$ (light grey).}}
\label{Fig-2}
\end{figure}
As is readily observed from Fig.\,\ref{Fig-2} the partial pressure $P_n$ exhibits 
asymptotic behavior. That is, the sum over $P_n$ seems to converge up to a (temperature dependent) 
critical value $n_c(T)$ but diverges when including 
contributions with $n>n_c(T)$. The asymptotic nature of the 
expansion of $P$ into $\sum_{n}P_n$ signals that an assumption made in deriving 
$P$ fails to hold for $n>n_c(T)$. Clearly, the assumption that up to arbitrarily large $n$ 
the associated solitons are stable and of mass $m_n=n\cdot\,m_e$ breaks down due to increasingly 
efficient contact interactions and annihilations. The decay or annihilation of large-$n$ 
solitons, however, modifies the spectral properties of small-$n$ 
excitations (increasing mass because of internal vibration and rotation) 
in such a way that the sum over partial pressures is likely to converge. 
The situation is somewhat reminiscent of Hilbert's hotel story 
where an unpleasant situation occuring at a finite integer $n$ 
is resolved by pushing this integer to infinity. Namely, 
to accommodate a naked, small-$n$ excitation in the spectrum 
(win of energy over entropy in the partition function) 
it needs to be dressed by additional internal energy which is released by the decay or annihilation 
of instable large-$n$ excitations. By induction this pushes $n_c(T)$ to infinity thus curing 
the apparent divergence of $\sum_{n}P_n$. For temperatures $T\le 0.6\,m_e$ 
it will be sufficient to restrict ourselves to 
$n\le 6$ where the expansion in terms of {\sl naked} excitations 
shows asymptotic convergence. A similar behavior is observed for the 
expansion of the energy density. 

\section{Explosive Z pinch}

Let us first lay out the scenario and point to experimental benchmarks 
for the Z pinch dynamics observed at Sandia. 
A wire array is fed with a current of about 
20\,MA. By resistive heating the wire is transformed 
into a plasma column. At the same time the current builds up 
a strong magnetic field whose pressure is directed inwards (towards the 
plasma axis): The plasma column, whose maximal radius is about 5\,mm, implodes until it 
stabilizes for about 5\,ns (stagnation). Shortly before, at, and shortly after 
stagnation soft X-ray emission is detected allowing for an 
estimate of the electron temperature $T_e\sim 3\,$keV. For the plasma 
implosion to stagnate the outward directed 
plasma pressure $P_{p}$ needs to be equal in magnitude to the inward directed 
magnetic pressure $P_{m}$ which is given as
\eqb
\label{P_m}
P_m=-\frac{1}{2}\,\frac{I_p^2}{\pi R_s^2}\,
\eqe
where $I_p$ is the plasma current, and 
$R_s$ denotes the radius of the plasma 
column at stagnation. Typically, one has 
$R_s\sim 1.5\,$mm and $I_p\sim 20\,$MA. This amounts to 
$P_m=-1.8\times 10^{-12}\,$MeV$^4$. For the electron density 
$n_e$ we take $n_e=Z\,n_i$ with $Z=26$ and the ion density 
$n_i=10^{26}\,\mbox{m}^{-3}$ \cite{Haines2006}. 
If one asserts that the plasma pressure $P_{p}$ 
is exclusively carried by electrons then the equilibrium 
condition $P_{p}+P_{m}=0$ predicts an electron 
temperature at stagnation of $T_e\sim 31.55\,$keV with an electron chemical 
potential $\mu_e=11.4\,$keV. Notice that $T_e$ 
is too high by a factor of $8.5$ in comparison with the observed 
value $T_e=3.6\,$keV. Thus we conclude that ions play a substantial 
role in the pressure balance at stagnation. This point was 
made very explicit in \cite{Haines2006}.  

According to 
\cite{Haines2006} the magnetic-field energy prevailing at 
stagnation is converted into ion heat by 
$m=0$ interchange MHD instabilities. This viscous heating 
mechanism increases the ion temperature 
to about $T_i\sim 300\,$keV very rapidly (time scale set by 
Alfv\'{e}n transit time $\tau_A\sim 1\cdots 2$\,ns). The increase of $T_i$ 
can be measured by an analysis of the Doppler broadening 
of the X-ray lines emitted by electron 
capture shortly after stagnation. Since the thermal 
equilibration time $\tau_e\sim 5\,$ns 
for electrons is significantly larger 
than $\tau_A$ and since the ion-ion 
collision time $\tau_{ii}$ is only 
$\tau_{ii}\sim 37\,ps$, $T_i$ will at first greatly 
exceed $T_e$. 

We are interested in what happens once a time 
$\tau_e\sim 5\,$ns has elapsed after stagnation. 
Ions will then have started to heat the 
electrons to $T_e\sim T_i\sim 300\,$keV at 
least locally. According to our discussion 
in the Sec.\,\ref{conf} this will involve center-vortex 
loops with a higher number of selfintersections. As a consequence, 
the equilibration time should decrease dramatically as compared to 
the conventional electron-gas picture: The electron 
temperature $T_e$ rises very rapidly at about $\tau_e\sim 5\,$ns after stagnation. 
In a conventional electron-photon plasma the Debye screening mass $m_D$ is 
given as
\eqb
\label{debayemass}
m_D\sim \sqrt{\frac{2}{T_e}}\,e\,\left(\frac{m_e\,T_e}{2\pi}\right)^{3/4}\,\exp(-m_e/(2\,T_e))
\eqe
where $e=\sqrt{4\pi\,\alpha}\sim 1/3$ the electromagnetic coupling. This represents a 
lower bound for the photon's electric screening mass being generated 
in the plasma discussed in Sec.\,\ref{conf}. Due to $T_e$ being a sizable fraction 
of $m_e$ after the ion-induced heating $m_D$ is, according to Eq.\,(\ref{debayemass}), 
comparable to $T_e$. This means that starting at about 5\,ns after stagnation 
no radiation is released by the then absolutely opaque 
plasma. At the same time, the plasma pressure will increase dramatically due to the presence 
of a large number of excitations with higher 
selfintersections. 

Let us be more quantitative about this. Define the truncated sums for the pressure and for the 
energy density in the electronic system as 
$\bar{P}_N=\sum_{n=1}^{N}P_{n}$ and $\bar{\rho}_N=\sum_{n=1}^{N}\rho _{n}$, 
respectively. For $N=6$ one is within the regime of asymptotic 
convergence for electron temperatures $<0.6\,m_e$. 
To make contact with the Z pinch experiment at Sandia 
we a priori would need to include a (small) chemical potential $\mu_e$ 
for preexisting electrons 
in the expression $\bar{P}_1$. However, we have checked that for $T_e>30\,$keV the chemical 
potential can safely be neglected. 

\begin{figure}[tbp]
\begin{center}
\leavevmode
\leavevmode
\vspace{4.3cm} \includegraphics{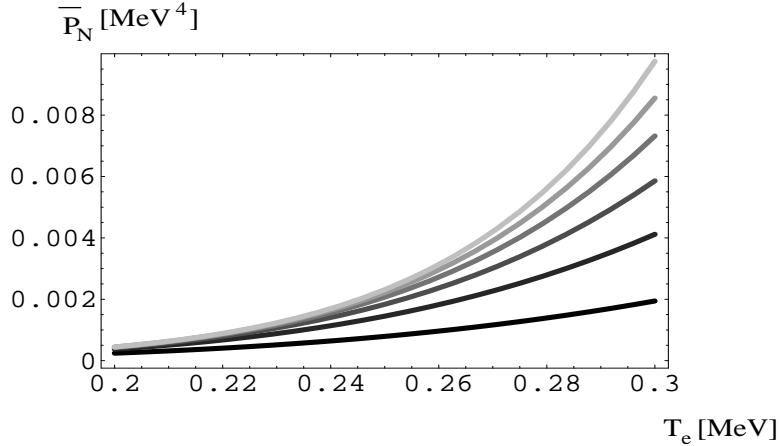}
\end{center}
\caption{{Plot of the truncated sums over partial pressures up to $N=6$ 
(dark grey: $N=1$; very light grey: $N=6$) as functions of electron temperature $T_e$.}}
\label{Fig-3}
\end{figure}

\begin{figure}[tbp]
\begin{center}
\leavevmode
\leavevmode
\vspace{4.3cm} \includegraphics{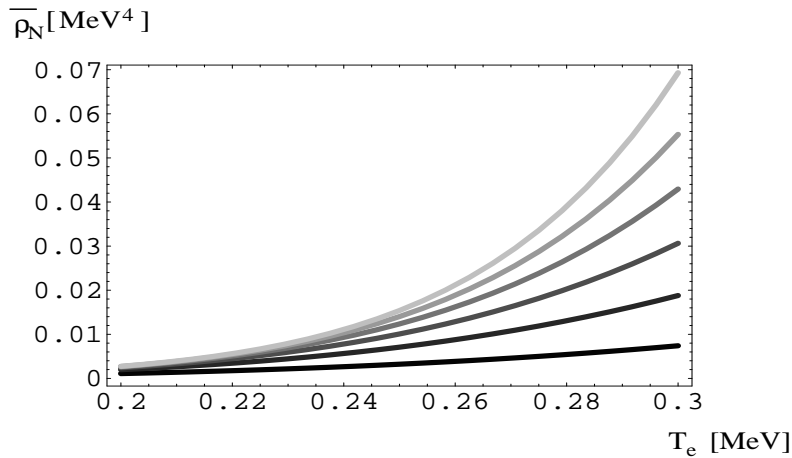}
\end{center}
\caption{{Plot of the truncated sums over partial energy densities up to $N=6$ 
(dark grey: $N=1$; very light grey: $N=6$) as functions of electron temperature $T_e$..}}
\label{Fig-4}
\end{figure}

In Figs.\,\ref{Fig-3} and \ref{Fig-4} we present plots of $\bar{P}_N$ and $\bar{\rho}_N$ 
in dependence of $T_e$ for $N=1,\cdots,6$.
While for temperatures $T_e\ll m_{e}$ no difference is visible (center-vortex loops with higher mass 
do practically not contribute) there is a clear enhancement of $\bar{P}_{6}$ and 
$\bar{\rho}_{6}$ with respect to $\bar{P}_1$ and $\bar{\rho}_1$ 
(contribution of electrons and positrons only) for $T_e\gtrsim 0.1$ MeV already. 
The deviation from $\bar{P}_1$ and $\bar{\rho}_1$ keeps growing with increasing $T_e$. 
We consider $T_e\sim 0.3$ MeV as an upper limit for the validity of 
our approximation (undressed and stable excitations). Going to
higher temperatures would require the precise knowledge of $M_{n>6}$, see Eq.\,(\ref{mn}), and
also would necessitate a consideration of finite widths and modified masses 
of the excited states in the evaluation of the partition function.

At $T_e=0.25$\,MeV we have $\bar{P}_6\sim 3\,\bar{P}_1$ and $\bar{\rho}_6\sim 4.8\,\bar{\rho}_1$, 
at $T_e=0.3$ MeV we already have $\bar{P}_6\sim 5\,\bar{P}_1$ and $\bar{\rho}_6\sim 9.4\,\bar{\rho}_1$. 
Notice that at $T_e=0.25$\,MeV the ratio of $\bar{P}_1$ to the magnetic pressure $P_m$ 
at stagnation is: $\frac{\bar{P}_1}{P_m}\sim -4.4\times 10^8$. We propose that 
$T_e$ reaching a value $\sim T_i\sim 0.3$\,MeV is facilitated by the very 
existence of center-vortex loops with higher mass. This would initiate 
the final stage in the Z-pinch dynamics leading to explosion. 

\section{Conclusion}

We have proposed a reason for the unexpected contained 
explosion of a Z pinch recently observed at Sandia 
National Laboratories. According to our scenario the presence 
of higher-mass center-vortex loops in the confining phase 
of SU(2)$_e$ accelerates the transit of thermal energy from 
ions to electrons and generates a larger pressure and energy density 
than expected from electron dynamics only. Once the electron 
temperature reaches a value of about $0.5\,$MeV a phase 
transition is expected to take place where all charged states 
condense into a new ground state (preconfining phase, ground state is superconducting \cite{Hofmann2005}). 
In that phase the dual gauge boson 
(identified with the $Z_0$ vector boson of the Standard Model) 
albeit massive propagates \cite{Hofmann2005,Hofmann2006,GiacosaHofmann2005}.  

On a qualitative level and more microscopically, we also predict that 
an extremely large (charge nonconserving) electron-positron multiplicity in the final state 
will be detected in hadron collisions at 
the LHC once the center-of-mass energy 
substantially exceeds the $Z_0$ mass, 
see also \cite{GiacosaHofmann2005}.

\section*{Acknowledgements}

One of us (F.G) would like to acknowledge financial support by G.S.I.

\baselineskip25pt

\end{document}